\documentclass[10pt,conference]{IEEEtran}
\usepackage{amsmath}
\usepackage{algorithmic}
\usepackage{graphicx}
\usepackage{textcomp}
\usepackage{xcolor}
\usepackage{listings}
\usepackage{framed}
\usepackage{stfloats}
\usepackage{tabularx}
\usepackage[table]{xcolor}
\usepackage{array}
\usepackage[hidelinks]{hyperref}
\usepackage{balance}  
\usepackage{cite}
\usepackage{caption}

\usepackage[most]{tcolorbox}
\definecolor{hdr}{HTML}{1C2541}
\definecolor{boxgray}{HTML}{EEF0F3}
\definecolor{accentA}{HTML}{2C4770}
\definecolor{accentB}{HTML}{2D5A27}
\definecolor{accentR}{HTML}{8B3A3A}

\newcolumntype{L}[1]{>{\raggedright\arraybackslash}p{#1}}

\newtcolorbox{keybox}[1][hdr]{colback=boxgray,colframe=#1,boxrule=0pt,leftrule=3.5pt,
  arc=1pt,left=6pt,right=6pt,top=4pt,bottom=4pt,boxsep=2pt}

\begin{document}

\title{ Is Model Instability  
 just Noise to be Tolerated\\ or a Property that can be Managed?}


\author{\IEEEauthorblockN{Amirali Rayegan}
\IEEEauthorblockA{\textit{North Carolina State University} \\
Raleigh, North Carolina, USA \\
arayega@ncsu.edu}
\and
\IEEEauthorblockN{Lunxiao Li}
\IEEEauthorblockA{\textit{North Carolina State University} \\
Raleigh, North Carolina, USA \\
lli66@ncsu.edu}
\and
\IEEEauthorblockN{Tim Menzies}
\IEEEauthorblockA{\textit{North Carolina State University} \\
Raleigh, North Carolina, USA \\
tjmenzie@ncsu.edu}
}

   \maketitle

\begin{abstract}
In software analytics, rerunning the same analysis twice often yields different models and conclusions. This reduces trust in the model and limits its use. We find that model instability is a major problem. Across 127 multi-objective SE optimization problems (12{,}700 test cases), repeated runs of a state-of-the-art optimizer agree on only \textbf{13.7\%} of test cases, even under improved settings. We argue that this instability is not merely noise to tolerate, but a property that can be measured and managed. By adjusting how labels are spent, how complex the models become, and how splits are scored, we obtain models that agree \textbf{4.8 times} as often as the default configuration. The standard deviation of optimization error falls by \textbf{22\%} on average (mean std 17.4 to 13.6), while recommendation quality improves rather than degrades. In terms of quality, the refined settings are statistically top-ranked on \textbf{119 of 127} datasets, compared to \textbf{74} for the defaults. We then test causal and data-locality interventions and find that they help only partially, suggesting a residual stability floor. Our evidence suggests there are fundamental limits to stability set by the data itself (noise, scarce labels, proxy objectives, and the many near-equivalent models a dataset admits). We conclude that instability should be treated as a standard evaluation axis in SE optimization, which should be routinely measured, reported alongside performance, and used to calibrate trust in any single run. The methods in this paper provide a baseline against which future efforts to reduce SBSE instability can be judged.

To support open science, we offer the following reproduction package:
\url{https://tinyurl.com/Model-Instability}
\end{abstract}


\section{Introduction}
\pagestyle{plain}  
\label{sec:introduction}
Software analytics is an influential research area with much practical value~\cite{abdellatif2015software, begel2014analyze, caldeira2023software, kotti2022impact, zhang2013software, he2022empirical}. By mining historical project data, teams allocate testing effort, prioritize high-risk modules, and cut maintenance costs. Eventually, though, stakeholders ask a deceptively simple question: ``What did you learn from all that data?'' The answer is usually a symbolic explanation, a decision tree, a rule set, or a causal graph. Unfortunately, these explanations are often \emph{unstable}. Rerunning the same analysis yields different models and, therefore, different conclusions. This reduces trust in the model and, therefore, limits its use.

For example, Figure~\ref{fig:instability_example} shows model variability seen after running the same optimizer several times on the same data using different random number seeds. As seen in Figure~\ref{fig:instability_example}, the same learner can produce models with different structures and use different attributes. The problem of model instability is not just a quirk of the learner used in  Figure~\ref{fig:instability_example}. Rather, it is endemic.  As SE problems grow larger and larger, state-of-the-art methods inject randomness to scale, and conclusion instability has been reported across regression~\cite{menzies2012local}, text mining~\cite{agrawal2018wrong}, defect prediction~\cite{bangash2020time}, causal discovery~\cite{hulse2025shaky}, and LLMs~\cite{er2025analyzing}.

\begin{figure*}[!t]
\centering
  \includegraphics[width=.85\textwidth]{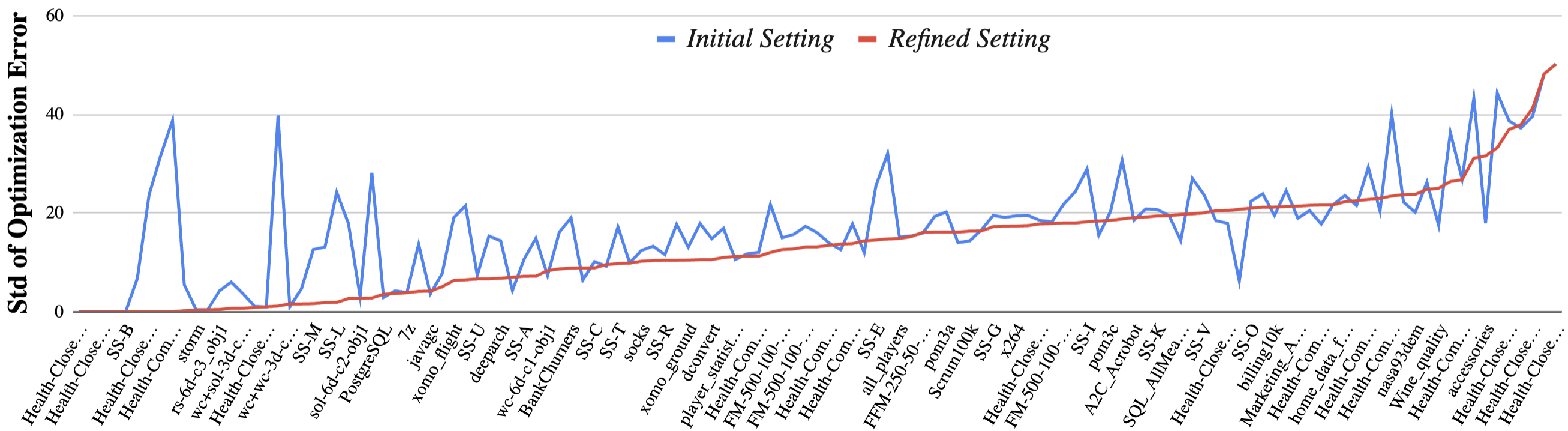}
  \caption{\textbf{Performance instability} across 127 SE datasets. Standard deviation of prediction errors 
  as defined by Equation~\ref{eq:win} (difference between the referenced optimal and the outcome of the model). Blue and red lines show, respectively, results obtained using (1)~the initial default settings or (2)~the settings recommended
  by this paper. Note that, with our methods (the red line), performance instability is nearly always reduced, sometimes by very large amounts (e.g., see left-hand side).}
  \label{fig:std}
  \vspace{-5pt}
\end{figure*}

\begin{figure*}[!t]
\centering
{\fontsize{6}{7}\selectfont
\begin{framed}
\vspace{-10pt}
\begin{lstlisting}
if TIME <= 4             >       if CPLx <= 4                  >      if PCAP > 1                    >       if DATA > 3              
|  if PCON > 3;          >       |  if PREC <= 5               >      |  if DOCU <= 2                >       |  if RUSE <= 3;         
|  if PCON <= 3          >       |  |  if FLEx > 5;            >      |  |  if STOR > 5;             >       |  if RUSE > 3;          
|  |  if ARCH <= 3       >       |  |  if FLEx <= 5            >      |  |  if STOR <= 5             >       if DATA <= 3             
|  |  |  if LTEx > 2;    >       |  |  |  if PCON <= 2         >      |  |  |  if DOCU > 1           >       |  if PCON > 1           
|  |  |  if LTEx <= 2;   >       |  |  |  |  if SITE > 2;      >      |  |  |  |  if PREC > 3;       >       |  |  if DATA > 2;       
|  |  if ARCH > 3;       >       |  |  |  |  if SITE <= 2;     >      |  |  |  |  if PREC <= 3;      >       |  |  if DATA <= 2       
if TIME > 4              >       |  |  |  if PCON > 2;         >      |  |  |  if DOCU <= 1;         >       |  |  |  if DOCU > 4;    
|  if SCED > 3;          >       |  if PREC > 5;               >      |  if DOCU > 2                 >       |  |  |  if DOCU <= 4    
|  if SCED <= 3          >       if CPLx > 4                   >      |  |  if SITE <= 3             >       |  |  |  |  if TIME > 4; 
|  |  if PMAT > 3;       >       |  if PMAT <= 1;              >      |  |  |  if TEAM > 3;          >       |  |  |  |  if TIME <= 4;
|  |  if PMAT <= 3;      >       |  if PMAT > 1;               >      |  |  |  if TEAM <= 3;         >       |  if PCON <= 1;         
                         >                                     >      |  |  if SITE > 3;             >                                 
                         >                                     >      if PCAP <= 1;                  >                               
\end{lstlisting}
\vspace{-10pt}
\end{framed}}
\vspace{-8pt}
\caption{\textbf{Structural instability.}
Example of models learned in this paper. Results are from 4 repeats varying only the random seed, extracted from \cite{rayegan2026can}, on Coc1000 dataset from MOOT (see Table~\ref{tab:moot_datasets}).}
\label{fig:instability_example}
\vspace{-10pt}
\end{figure*}

A natural reaction is to treat instability as a bug to fix with more data, tuning, or ensembling. But the literature suggests a harder truth. Often, there is no single ``best'' explanation to recover. Breiman's ``two cultures'' essay notes that structurally different models can fit the same data nearly as well~\cite{breiman2001statistical}, the \textit{Rashomon effect}, which Xin et al. formalize as the (often enormous) \textit{Rashomon set} of near-optimal models for a task~\cite{xin2022exploring}. When many models are nearly tied, small perturbations steer learning toward different structures. Structural instability is thus not a defect to eliminate but a property of the data.

This reframes what ``stability'' should mean. We distinguish \emph{performance instability} (variation in the outcomes of recommended solutions, see Figure~\ref{fig:std}) from \emph{structural instability} (variation in model form and features used, see Figure~\ref{fig:instability_example}). Expecting the same model every time is the wrong goal. What matters is whether the analysis yields consistent \emph{recommendations}.  We therefore assert that performance instability is the pressing problem we target. Based on that, this paper asks four research questions.
\begin{enumerate}
    \item \textbf{RQ0 -- The Problem:} How prevalent is performance instability in SE optimization, and what are its consequences for practitioners?
    \item \textbf{RQ1 -- The Nature:} How prevalent is structural instability, and is it the same as performance instability?
    \item \textbf{RQ2 -- The Fixes:} Which factors drive performance instability, and what configuration changes reduce it without sacrificing optimization quality?
    \item \textbf{RQ3 -- The Limits:}  Is the instability remaining after RQ2 the fault of the learner, or of the data?
\end{enumerate}
This paper contributes (1) a large-scale study of 127 SE optimization problems showing pervasive instability (repeated runs agree on only 13.7\% of test cases under refined settings); (2) evidence that structural and performance instability are distinct. Trees can be structurally diverse yet yield consistent suggestions, making performance stability the only actionable criterion; (3) a systematic analysis identifying labeling budget, sampling strategy, model complexity, and splitting criterion as independent drivers, with a validated configuration that improves both stability and quality; and (4) evidence of a data-inherent floor that learner-side and causal interventions reduce but cannot fully eliminate. 

To support open source, all our data and scripts are available at \url{https://tinyurl.com/Model-Instability}.

Before we begin, we clarify how to read the main reported numbers and why they are important and new. We use three complementary summaries. First, test-case agreement measures performance stability. Out of 12{,}700 held-out test cases, how often do 20 repeated runs give sufficiently similar predictions? At our main threshold $\alpha=0.35$, agreement improves from 364 cases under defaults to 1{,}740 cases under refined settings, a 4.8$\times$ increase. Second, error spread measures the standard deviation of optimization error across repeated runs. As shown in Figure~\ref{fig:std}, the red line is mostly below the blue line; refined settings reduce the mean standard deviation of errors from 17.4 to 13.6, a 22\% reduction across all 127 datasets. While Figure~\ref{fig:std} asks whether repeated runs are more stable, Figure~\ref{fig:rq2_final_comparison} asks whether the resulting recommendations remain high quality. Moving to the third, optimization quality is assessed by statistical top-rank counts across datasets. Refined settings are top-ranked on 119 of 127 datasets, compared to 74 for defaults. Thus, agreement and error spread describe stability, while top-rank counts describe recommendation quality. Importantly, the stability gains come with no performance compromise. These results matter because, although instability has been reported in several areas of software analytics, it is still rarely treated as a first-class evaluation target. Prior work has shown that learned conclusions can vary across data contexts in defect prediction and effort estimation~\cite{menzies2012local}, that defect predictors can differ in robustness and stability~\cite{kaur2015empirical}, that bellwether effects complicate transfer learning across projects~\cite{krishna2018bellwethers}, that deep-learning software systems exhibit substantial training variance and reproducibility challenges~\cite{pham2020problems,10.1145/3477535}, and that causal graphs learned from SE data can be structurally unstable~\cite{hulse2025shaky}. Yet studies typically emphasize mean optimization performance, while reporting little about the variance of the recommendations produced by repeated runs. To the best of our knowledge, this paper is the first large-scale study of model instability in SBSE (``large scale'' since, as shown below, we report results from 127 SE problems and 12{,}700 test cases). Stabilizing models at that scale shows that instability is not merely noise to tolerate, but a property that can be measured and managed. The methods presented here should therefore be viewed as a baseline for future work on SBSE instability.

\section{Background}
\label{sec:background}
\subsection{Motivation}
Why study instability in SE optimization? Because SE systems rarely optimize a single objective. Tasks such as configuration tuning, test prioritization, and project scheduling must balance competing goals, including:
\begin{enumerate}
\item Run {\em more} database queries, {\em faster},
      with {\em less} energy;
\item Deliver {\em better} software, {\em faster}, yet {\em cheaper};
\item {\em Maximize} stakeholder satisfaction with {\em less} complexity; 
\item Test for {\em more} bugs with {\em less} effort.
\end{enumerate}

Search-based software engineering exists to negotiate exactly these trade-offs. The choice spaces it must explore are enormous. Configuration spaces are combinatorially vast\footnote{For example, Table 1 of Agrawal et al.~\cite{agrawal2019dodge} reports one hundred sextillion ($10^{46}$) ways to configure text-preprocessing data miners.}, under-documented, and laced with subtle interactions. Defaults cannot be trusted, hand-tuning does not scale, and human intuition often misleads~\cite{easterby1980design}. As systems grow more intricate, managing their configurations becomes increasingly difficult~\cite {xu2015hey}. To navigate vast search spaces, learning algorithms often use random search, causing repeated runs to yield different models and recommendations, leading to structural and performance instability.

\begin{figure}[!t]
\centering
  \includegraphics[width=.9\columnwidth]{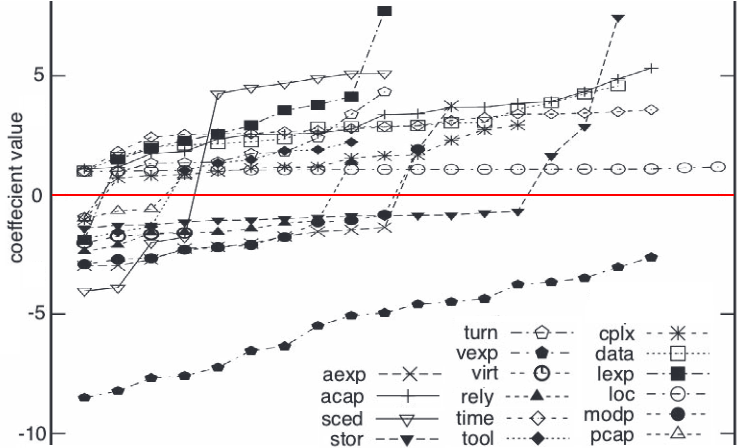}
  \caption{Coefficient instability in effort estimation~\cite{menzies2012local}. $\beta_i$ coefficients from 20 samples of the same dataset, with several crossing zero thresholds, showing that same feature increases effort in some samples, decreases it in others.
}
  \label{fig:lr-coeff}
  \vspace{-10pt}
\end{figure}

This instability has been seen in many SE domains. In \textbf{software project effort estimation}, regression learns \mbox{$y = \beta_0 + \sum_i \beta_i x_i$}, where positive or negative $\beta_i$ means feature $x_i$ increases or decreases estimated effort. In theory, those $\beta_i$ could be used to prioritize changes within a project. In practice, instability makes this very difficult.  Figure~\ref{fig:lr-coeff}, from Menzies et al.~\cite{menzies2012local}, shows the $\beta_i$ learned from 20 different 66\% samples of the same project dataset. The coefficients swing wildly, and half a dozen even cross $\beta_i = 0$. The same feature is estimated to both \textit{increase} and \textit{decrease} effort, making any consistent interpretation impossible for practitioners\footnote{Even a simple regression can fluctuate so wildly. See all the parameters in scikit-learn's LinearRegression: \url{https://scikit-learn.org/stable/modules/generated/sklearn.linear_model.LinearRegression.html}.}.

In SE \textbf{topic modeling}, Agrawal et al.~\cite{agrawal2018wrong} show
that standard LDA (Latent Dirichlet allocation) is severely unstable. Prior
researchers had used LDA to report ``what developers talk about on Stack
Overflow,'' with no hint that their topics were unstable. That was
misleading. Merely randomizing the input order dramatically changes the
topics reported. Running LDA twice with different orderings
(Table~\ref{tab:lda-instability-example}), only 3 of 40 topics recurred, and over half (25 of 40) had 55\% or less overlap.

At the \textbf{conclusion level in defect prediction}, Bangash et
al.~\cite{bangash2020time} found that conclusions about model performance
drift over time. Across five cross-project defect prediction techniques and
33 versions of 14 Java projects, rankings and performance shifted by
statistically significant margins in time-aware evaluations, driven by
time, but also by dataset noise and the nature of CPDP approaches.

For \textbf{causal graphs}, widely used as succinct representations of complex SE data, Hulse et al.~\cite{hulse2025shaky} report highly unstable shapes. When applying four generators to 23 datasets across three SE tasks, over half the causal edges disappeared or changed direction. Around 75\% of edges differ between consecutive project releases, and removing just 10\% of training data can cause large structural changes. Any specific causal conclusion (e.g. ``long functions cause more defects'') could be reversed by minor changes to data or method.

With the growing adoption of \textbf{LLMs in SE}, the same concern recurs. A seminar of SE/AI researchers flagged the instability of generated results as a core obstacle to reliable LLM4SE~\cite{gao2025current}, and the effect is empirically reproducible. In error correction, LLMs are demonstrably unstable, with fix consistency degrading as the sampling temperature rises~\cite{er2025analyzing}. In vulnerability detection, Han et al.~\cite{han2025prompting} tested 6 LLMs on 48 scenarios drawn from the MITRE Top-25 weaknesses, across 17 prompt structures with 20 paraphrases each. Semantically equivalent rephrasings alone produced substantially more output variability than repeated identical prompts, with code-specialized models more brittle than general-purpose ones.

In summary, instability is not limited to specific techniques or tasks. It
is pervasive across SE research, observed in \textbf{reproducibility}
studies~\cite{semmelrock2025reproducibility} and \textbf{large-scale
empirical analyses}~\cite{hoess2025oopsididagain}. Even methodological
choices alone,\textbf{tuning}~\cite{fu2016tuning} or \textbf{validation
strategy}~\cite{ali2021empirical},can shift rankings and conclusions.
Hence, we say instability is a first-class, systemic issue in software analytics and must be actively investigated and mitigated.

\begin{table}[t]
\scriptsize
\renewcommand{\arraystretch}{1.2}
\begin{tabular}{ccc}
\textbf{\% repeats in run 2} & \textbf{Topic name} & \textbf{Top 9 words in topic} \\
\hline
100 & Xaml Binding       & grid,window,bind, \ldots \\
100 & Ruby on Rails      & rubi,rail,gem, \ldots \\
100 & MVC                & http,com,java, \ldots \\
88  & Objective C        & self,cell,nsstring, \ldots \\
88  & Function Return Types & function,const,char,\ldots \\
\multicolumn{3}{c}{\ldots} \\
55  & MySQL              & connect,sql,databas, \ldots \\
55  & HTML Form          & form,login,user, \ldots \\
\multicolumn{3}{c}{\ldots} \\
22  & Visual Studio      & window,project,file, \ldots \\
22  & Information Systems & messag,email,log, \ldots \\
\hline
\end{tabular}
\setlength{\abovecaptionskip}{4pt}
\setlength{\belowcaptionskip}{-10pt}
\caption{Examples of LDA topic instability across two runs, adapted from Agrawal et al.~\cite{agrawal2018wrong}.}
\label{tab:lda-instability-example}

\end{table}

\subsection{Structural Instability and the Rashomon Effect}
\label{sec:rashomon}
This paper measures both structural and performance instability but seeks to mitigate only the latter. The reason is not neglect. Theory predicts (and RQ1 confirms) that structural instability is fundamental, an unavoidable consequence of the many different models that fit the same data.

Structural instability is a variation in model form and/or the features used. To understand its nature, consider a model with $M=10$ binary inputs: $2^{10}=10^{24}$ input patterns, of which the model may internally use any subset. There are \[2^{2^{10}} \approx 1.8 \times 10^{308}\] such subsets (more than the electrons in the observable universe). This is the Rashomon effect: an enormous number of structurally different models can fit the same data~\cite{breiman2001statistical,xin2022exploring}.

Semenova et al.~\cite{semenova2022simpler} formalized this via the
\textit{Rashomon ratio}: the fraction of models achieving near-optimal
performance. When that ratio is large, the selected model reflects random
variation in the data, not a stable underlying signal. Rudin et al.\ showed
the Rashomon set can be enormous for tabular data~\cite{rudin2024amazing}.
Xin et al.~\cite{xin2022exploring} showed that for sparse decision trees, it
can be exponentially large, making structural variability the norm.
Crucially, Semenova et al.~\cite{semenova2023path} explain \textit{why}
Rashomon ratios tend to be large. Label noise widens generalization gaps
and forces simpler model classes with greater multiplicity. SE datasets,
often noisy~\cite{wu2021data, Seiffert2007noisy}, small, and labeled under
tight budgets~\cite{8469102, RAYEGAN2026112897}, are precisely those
conditions. From this perspective, structural instability in SE
optimization is not an algorithmic defect; it is a \textbf{property of the
problem itself}.

In summary, theory predicts structural instability is something to {\em
measure}, not {\em fix}. RQ1 confirms this empirically. Even settings that
greatly improve prediction agreement leave the tree structure as diverse. Hence, the rest of this paper measures both instabilities but only
mitigates performance instability.

\subsection{Causes of Instability}
\label{sec:causes}
The Rashomon effect implies that the same data can support many models. A learner's inductive biases determine which one is selected. Those same biases also create instability: change a design choice (e.g., the random seed), and the learner may produce a different model. There are many such choices. This section examines six, summarized in Table~\ref{tab:instability_causes}, ranging from familiar tuning knobs to subtler effects. The first four are internal to the learner (RQ2), and the last two concern the learner's interaction with the data (RQ3).

No such list is complete. We chose these six because they are widely applicable (most affect a broad range of learners), effective (as shown below, they can lift stability from under 50\% to over 80\%), and vetted (some (e.g., Causal-inspired) were added after discussion with leading figures in the empirical SE community. Other candidates were ruled out experimentally: feature selection, initially included, in fact {\em increased} instability, since with fewer features the variance of a single attribute can dominate.

\textbf{Labeling Budget.} A learner builds a model $y=f(x)$ from examples $(x_1,y_1), (x_2,y_2),...$. We denote by $ y_i$ the label of each example. Obtaining labels is the costly part of SE optimization since it means running or deploying a real system\footnote{Finding authoritative labels remains a major challenge for SE research~\cite{ahmed2025can,h2022}. Labeling by human experts is slow and error-prone when rushed~\cite{easterby1980design,valerdi2010heuristics}, historical logs can be unreliable~\cite{yu2020identifying,wu2021data,kang2022detecting,shepperd2013data}, and automated labeling is crude~\cite{kamei2012large} or, in the case of LLMs, only assistive (not authoritative). Even naturally occurring oracles (compile, then run the full test suite) can be extremely slow~\cite{DBLP:conf/wosp/ValovPGFC17}.}. The {\em labeling budget} caps how much data the learner ever sees, so with small budgets, each draw sees a different set of rows and tells a different story. We sweep budgets from 10 to 200 (upper bound is empirically justified in \S\ref{sec:exp-datasets}). 

\textbf{Model Complexity.} A shallow model yields a few coarse rules, while a deeper one chases nuances supported by less and less data, so sample size can change a tree's lower structure. In decision tree learning, the minimum leaf size controls the model's complexity.

\begin{table}[!t]
\centering
\begin{tabularx}{\columnwidth}{
>{\hsize=0.03\hsize}X
>{\hsize=0.7\hsize}X
>{\hsize=1.4\hsize}X
c
}
\textbf{\#} & \textbf{Instability Cause} &
\textbf{Treatments in Experiment} & \textbf{RQ} \\
\hline
1 & Labeling Budget
  & Labels = [10, 20, 50, 100, 200]
  & 2 \\
2 & Acquisition \newline Strategy
  & [Xplor, Xploit, adapt, bore, \newline near, random]
  & 2 \\
3 & Model Complexity
  & Min\_leaf size = [1, 3, 5, 7, 9]
  & 2 \\
4 & Splitting Criterion
  & [Entropy (uses log), Gini (no log)]
  & 2 \\
5 & Causal-inspired
  & Confounder filter \& Gain-Ratio split = [On,Off]
  & 3 \\
6 & Data Locality
  & Anchors = [HDBSCAN, CURE, KMeans]
  & 3 \\
\hline
\end{tabularx}
\vspace{2px}
\setlength{\abovecaptionskip}{0pt}
\setlength{\belowcaptionskip}{-10pt}
\caption{Potential sources of performance instability and the corresponding
experimental treatments. The first four are studied in RQ2, the last two in
RQ3.}
\label{tab:instability_causes}
\vspace{-5pt}
\end{table}

\textbf{Splitting Criterion.} Models divide regions of data in order to, say, minimize expected post-division impurity. There are many measures of impurity. Entropy counts the bits needed to encode a distribution via $\left(-\log p\right)$, which runs to infinity as $p \to 0$. Since its slope is $1/p$, small changes in rare-event probabilities can dramatically change what is selected by this particular splitting criterion. Other criteria are not so unstable. Gini impurity ($1-\sum_i p_i^2$) has a bounded slope of  $2p$, so rare events cannot dominate splits (the way they do with entropy).

\textbf{Acquisition Strategy.} {\em Active learners} use the model built so far to decide what row label to acquire next, thus avoiding irrelevant and noisy data. Different acquisition rules select different rows and yield different theories, making the rule itself a source of instability. After sorting labeled data into better and worse regions, our learners estimate the probability that an unlabeled row is best ($b$) or rest ($r$). From the literature, we extract five rules. \textit{Bore} and \textit{near} greedily seek the best-not-rest row via Bayes statistics (bore) or distance to the best centroid (near). \textit{Xplor}, \textit{xploit}, and \textit{adapt} label the row maximizing
\[
\frac{b+rq}{\mathit{abs}(bq - r + \epsilon)}
\]
where $\epsilon$ avoids divide-by-zero and
\begin{equation}\label{q}
q=\begin{cases}
0 & \text{if exploiting}\\
1 & \text{if exploring}\\
1 - \frac{M}{B} & \text{if adapting}
\end{cases}
\end{equation}
($M$ = configurations evaluated so far; $B$ = total budget). When labels are scarce, {\em explore} favors rows where the best and rest votes are close, opposite ideas at similar weight. As data accrue, {\em exploit} jumps to the strongest best and weakest rest; {\em adapt} slides from explore to exploit as $M$ grows toward $B$.

\begin{table*}[!t]
\centering
\scriptsize
\setlength{\tabcolsep}{4pt}
\begin{tabularx}{\textwidth}{
c
p{2.5cm}
p{5.5cm}
p{5cm}
c
c
}
\rowcolor{gray!20}
\textbf{\#} &
\textbf{Dataset Type} &
\textbf{File Names} &
\textbf{Primary Objective} &
\textbf{x/y} &
\textbf{\# Rows} \\
\hline

25 &   Software Configurations
& SS-A to SS-X, billing10k
& Optimize software system settings
& 3--88 / 2--3
& 197--86,059 \\

12 & PromiseTune \newline Software Configurations
& 7z, BDBC, HSQLDB, LLVM, PostgreSQL, dconvert, deeparch, exastencils, javagc, redis, storm, x264
& Software performance optimization
& 9--35 / 1
& 864--166,975 \\

1 & Cloud
& HSMGP num
& Hazardous Software Management Program data
& 14 / 1
& 3,457 \\

1 & Cloud
& Apache AllMeasurements
& Apache server performance optimization
& 9 / 1
& 192 \\

1 & Cloud
& SQL AllMeasurements
& SQL database tuning
& 39 / 1
& 4,654 \\

1 & Cloud
& X264 AllMeasurements
& Video encoding optimization
& 16 / 1
& 1,153 \\

7 & Cloud
& (rs--sol--wc)*
& Misc configuration tasks
& 3--6 / 1
& 196--3,840 \\

35 & Software Project Health
& Health-ClosedIssues, Health-PRs, Health-Commits
& Predict project health and developer activity
& 5 / 2--3
& 10,001 \\

3 & Scrum
& Scrum1k, Scrum10k, Scrum100k
& Configurations of the scrum feature model
& 124 / 3
& 1,001--100,001 \\

8 & Feature Models
& FFM-*, FM-*
& Optimize number of variables, constraints, \newline and clause/constraint ratio
& 128--1,044 / 3
& 10,001 \\

1 & Software Process Model
& nasa93dem
& Optimize effort, defects, time, and LOC
& 24 / 3
& 93 \\

1 & Software Process Model
& COC1000
& Optimize risk, effort, analyst experience, etc.
& 20 / 5
& 1,001 \\

4 & Software Process Model
& POM3 (A--D)
& Balance idle rates, completion rates, and cost
& 9 / 3
& 501--20,001 \\

4 & Software Process Model
& XOMO (Flight, Ground, OSP)
& Optimize risk, effort, defects, and time
& 27 / 4
& 10,001 \\

2 & Software Testing
& test120, test600
& Optimize test selection
& 9 / 1
& 5,161 \\

\hline

3 & Miscellaneous
& auto93, Car\_price, Wine\_quality
& Miscellaneous optimization tasks
& 5--38 / 2--5
& 205--1,600 \\

4 & Behavioral
& all\_players, HR-employeeAttrition, \newline student\_dropout, player\_statistics
& Analyze and predict behavioral patterns
& 26--55 / 1--3
& 82--17,738 \\

4 & Financial
& BankChurners, home\_data, Loan, Telco-Churn
& Financial analysis and prediction
& 19--77 / 2--5
& 1,460--20,000 \\

3 & Human Health Data
& COVID19, Life\_Expectancy, hospital\_Readmissions
& Health-related analysis and prediction
& 20--64 / 1--3
& 2,938--25,000 \\

5 & Sales
& accessories, dress-up, Marketing\_Analytics, \newline socks, wallpaper
& Sales analysis and prediction
& 14--31 / 1--8
& 247--2,206 \\

2 & Reinforcement Learning
& A2C\_Acrobot, A2C\_CartPole
& Reinforcement learning tasks
& 9--11 / 3--4
& 224--318 \\\hline

\textbf{127} & \textbf{Total} & & & & \\

\end{tabularx}
 \caption{Summary of the 127 datasets used in this study from  Chen \& Menzies' MOOT repository  \url{http://tiny.cc/moot}~\cite{menzies2026mootrepositorymultiobjectiveoptimization}. The table reports dataset categories, representative file names, optimization objectives, number of decision variables and objectives (x/y), and dataset sizes.}
\label{tab:moot_datasets}
\vspace{-10pt}
\end{table*}

\textbf{Data Locality.} SE repositories often combine data from different developers, technologies, and problem domains. When modeled as a single population, such heterogeneous data may support many equally plausible models (Rashomon again), increasing instability. One way to reduce this effect is to first partition the data into more homogeneous regions. To test whether stability gains depend on a particular clustering paradigm, we evaluate one method from each major family: \textbf{KMeans}~\cite{macqueen1967kmeans} (partitional clustering around iteratively updated centroids), \textbf{HDBSCAN}~\cite{campello2013hdbscan} (density-based clustering that discovers arbitrarily shaped regions), and \textbf{CURE}~\cite{guha1998cure} (representative-point clustering that captures non-spherical structures while remaining robust to outliers).

\textbf{Causal-inspired.} Pearl warns that learners built on raw correlation invent influences that do not exist~\cite{pearl2009causality, pearl2018book}, lacking causal knowledge, they cannot tell whether $A$ correlates to $B$ because $A$ drives $B$, or because both ride a hidden third variable. Such spurious inferences inflate the space of plausible models. Causal knowledge can steady a learner. Unicorn~\cite{iqbal2022unicorn} uses causality to sharpen configuration models, and PromiseTune~\cite{Chen2026PromiseTune} filters spurious predictors before search.

Following PromiseTune, we add a confounder filter to our model construction. Each labeled row gets a target $Y$ equal to its $d2h$ score (numerics discretized into 10 equal-probability bins). The dependency of feature $X$ on $Y$ is the drop in target impurity after conditioning on $X$:
\begin{equation}
    D(X, Y) = I(Y) - \sum_{x} p(x)\, I(Y \mid X = x)
\end{equation} 
where $I(\cdot)$ is entropy or Gini. Features are kept when $D(X,Y)$ beats a permutation null at $\alpha$ significance. Among survivors, if conditioning on some $Z$ renders $D(X, Y \mid Z)$ negligible, $X$ is removed as confounded by $Z$ (mutual confounders tie-break by smaller $H(Y \mid X)/H(Y)$). After filtering, splits are scored by the gain ratio $1 - H(Y \mid X)/H(Y)$, the most stable splitting criterion per Lee et al.~\cite{lee2024assessing}. (Aside: note our terminology. we use ``causal'' in the sense of confounder screening per PromiseTune~\cite{Chen2026PromiseTune} and not as per the full do-calculus.)

\section{Experimental Design}
\label{sec:experimental-design}
This section describes the data, measures, and algorithms used to answer our four research questions.

\subsection{Data}
\label{sec:exp-datasets} We use 127 multi-objective SE optimization problems from Chen \& Menzies' MOOT repository \url{http://tiny.cc/moot}~\cite{menzies2026mootrepositorymultiobjectiveoptimization}. See Table~\ref{tab:moot_datasets}. For our experiments, each  dataset is split randomly 50:50 into train and hold-out halves (and all our runs are from 20 such runs).
We use MOOT since, to our knowledge, it is the largest collection of real multi-objective SE optimization tasks assembled to date. MOOT's data come from numerous SE papers by numerous SE authors, presented in leading venues (ICSE, FSE, TSE, IST, EMSE, TOSEM, ASE). MOOT's tasks span configuration, performance tuning, product lines, project health, defect prediction, testing, cost estimation, cross-domain generalization, and text mining. Earlier resources (e.g., SPLOT) were narrow and are now offline. Toolkits like Pygmo or Platypus offer only synthetic benchmarks. MOOT's datasets come from published studies, real performance logs, cloud systems, and tuning tasks where bad configurations cost time, money, and credibility.

\begin{figure*}[!t]
  \centering
  \includegraphics[width=\textwidth]{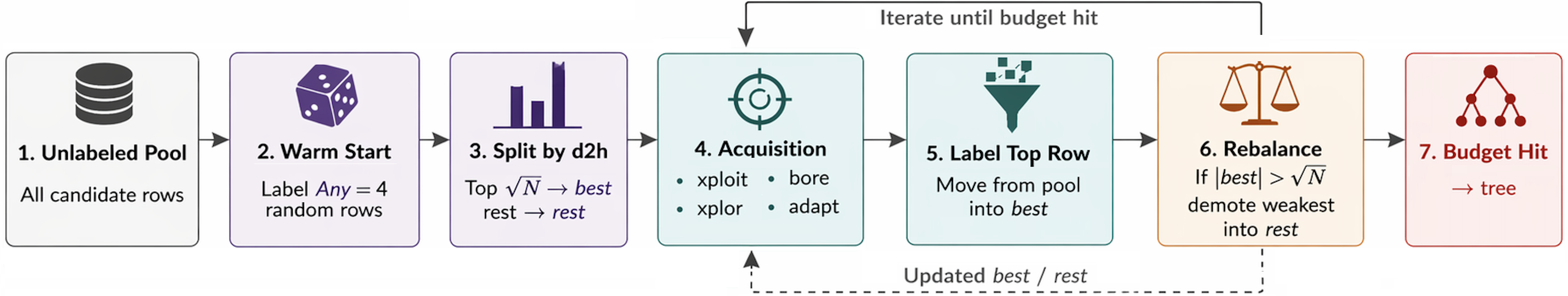}
  \caption{Overview of the EZR active-learning pipeline,
    from warm-start initialization through iterative
    acquisition and labeling to final tree construction.}
  \label{fig:ezr-pipeline}
  \vspace{-10pt}
\end{figure*}

One note on data access. While MOOT problems can hold hundreds to 100,000s of rows (see Table~\ref{tab:moot_datasets}), we demand that our optimizers access the $y$ goal values of no more than 200 rows. That cap was set empirically. We increased the labeling budget until further labels stopped buying performance, a ceiling reached at 200 labels (see Figure~\ref{fig:rq2_combined}, top-left). This is consistent with (indeed, slightly more generous than) the 150-label saturation effect reported by Menzies and Srinivasan~\cite{menzies2026aieasylessonslearned}, who found no significant improvements after 150 examples (where those examples were carefully selected by the optimizer to avoid noisy and irrelevant items).

\subsection{Measures}
\label{sec:measures}
Configuration quality is measured as  ``distance-to-heaven'' ($d2h$), normalized to {\em win} (to allow for comparisons across different data):

{\scriptsize \begin{equation}\label{eq:q}
d2h(x)=\sqrt{\sum_{i=1}^{m} (o_i(x) - h_i)^2},
\quad
Q=\mathit{d2h}(x) - \min,
\end{equation}
\begin{equation}
\text{win}(x) = 100\left(1 - \frac{Q}{\text{median} - \min}\right)
\label{eq:win}
\end{equation}}

Here, $o_i(x)$ is the normalized objective of configuration $x$, $h_i$ the ideal value of objective $i$, and $m$ the number of objectives. Also, $\min$ and {\em median} are the lowest and middle values for {\em d2h} seen in the data set. Note that  {\em smaller} $d2h$ and  $Q$ values are {\em better} while {\em larger} values for {\em wins} are  {\em better}. (Aside: later in this paper, we will refer to $Q$ as the {\em optimization quality}).

One definition is used for everything below: every algorithm in this study
returns a function $\mathit{Model}(\mathit{row}) \rightarrow$ predicted win
score. Recommendations, performance, and all stability measures are computed
from these predictions, so trees, clusterers, and baselines are scored on the
same footing.

\subsection{Primary Learner}
\label{sec:ezr}

Our primary learner is \textbf{EZR}~\cite{menzies2025case}, the active-learning optimizer shown in Figure~\ref{fig:ezr-pipeline}. Recent studies in 2026 by Rayegan et al. and Ganguly et al., published in JSS~\cite{RAYEGAN2026112897} and FSE~\cite{ganguly2025low}, establish EZR as a state-of-the-art baseline in SE multi-objective optimization~\cite{ganguly2026zoom} and explanation~\cite{RAYEGAN2026112897,menzies2026aieasylessonslearned}. EZR learns decision trees from only a few dozen labeled examples while achieving state-of-the-art optimization performance and running up to $500\times$ faster than methods such as SMAC3~\cite{lindauer2022smac3}.

We study EZR for three reasons. \textbf{First}, it is a recent state-of-the-art method. \textbf{Second}, it exhibits the instabilities investigated in this paper (all models in Figure~\ref{fig:instability_example} were generated by EZR). \textbf{Third}, it has already been evaluated on the 127 MOOT tasks described in \S\ref{sec:exp-datasets}, allowing us to assess stability across a large and diverse benchmark suite.

EZR begins by labeling a small random warm-start sample. These examples are ranked by $d2h$ and divided into \textit{best} and \textit{rest}. An acquisition function then iteratively selects additional rows for labeling, updating the \textit{best}/\textit{rest} partition after each new label, until the labeling budget $n_1$ is exhausted. We evaluate six acquisition functions (Table~\ref{tab:instability_causes}, formulas in \S\ref{sec:causes}).

Once labeling is complete, EZR constructs a regression tree from the $n_1$ labeled examples. Each split selects the feature and threshold that minimize the expected standard deviation of $d2h$ in the resulting partitions. Recursion continues until the minimum leaf size is reached. Each leaf predicts the median $d2h$ of its training examples, which is then converted to the win score defined in \S\ref{sec:measures}. To generate a recommendation, the learned model ranks the hold-out examples by predicted score. The top $n_2$ candidates are then labeled, and the best among them is returned. The entire process, therefore, consumes $n_1+n_2$ labels. Throughout this paper, we use the default value $n_2=10$, so all reported labeling budgets should be interpreted as requiring an additional ten labels.

{\em Important note:}
Throughout the results, we compare two EZR configurations:
\textbf{initial} (the defaults of~\cite{menzies2026aieasylessonslearned})
and \textbf{refined} (Table~\ref{tab:config_recommendation}). RQ0 and RQ1 use
both as fixed treatments; RQ2 reports the factor-by-factor experiments from
which the refined settings were derived.

\textbf{Other learners.} RQ3 also runs three clusterers (KMeans, HDBSCAN, CURE; rationale in \S\ref{sec:causes}) as stability anchors. Each implements $\mathit{Model}(\mathit{row})$ by returning the median $d2h$ of the labeled rows in the cluster nearest to that row. RQ3's causal variant applies the confounder filter and gain-ratio splits of \S\ref{sec:causes} to EZR's tree construction; nothing else changes.

\subsection{Measuring Performance}
\label{sec:perf_measure}
Each treatment is repeated 20 times with different random seeds. For run $i$, we compare the recommendation $\hat{x}_i$ against $x^*_i$, the true optimum (lowest $d2h$) in the hold-out:

{\scriptsize \begin{equation}
\text{RMSE} = \sqrt{\tfrac{1}{20}
\sum_{i=1}^{20} \left(
\text{win}(\hat{x}_i) - \text{win}(x^*_i)
\right)^2}
\end{equation}}
Note that {\em lower} RMSE is  {\em better}. Applying \S\ref{sec:statistics} to per-dataset RMSE yields each treatment's \textit{performance win count}: the number of datasets (out of 127) where each treatment has statistically better suggestions compared to other methods.

\subsection{Measuring Stability}
\label{sec:stab_measure}
The same 20 models per treatment per dataset support two stability measures: one structural, one performance-based.

\textbf{Structural stability} asks whether repeated runs produce similar model structures. Following Kalousis et al.~\cite{10.1007/s10115-006-0040-8}, we use Jaccard similarity to compare the structure of two trees trained under identical settings. However, since the features appearing in splits closer to the root of the tree gain more importance, we measure the weighted Jaccard similarity:
\begin{equation}
[J_w(A,B)=
\frac{
\sum\limits_{f \in F}
\min\left(W_A(f),W_B(f)\right)
}{
\sum\limits_{f \in F}
\max\left(W_A(f),W_B(f)\right)
}]
\end{equation}

where (F) is the union of features appearing in either tree, and ($W_T(f)=\sum_{k \in S_f} 1/d_k$) assigns larger weights to features occurring closer to the root. The similarity ranges from 0 (disjoint features) to 1 (identical weighted features).

\begin{figure}[!b]
\centering
\includegraphics[width=\columnwidth]{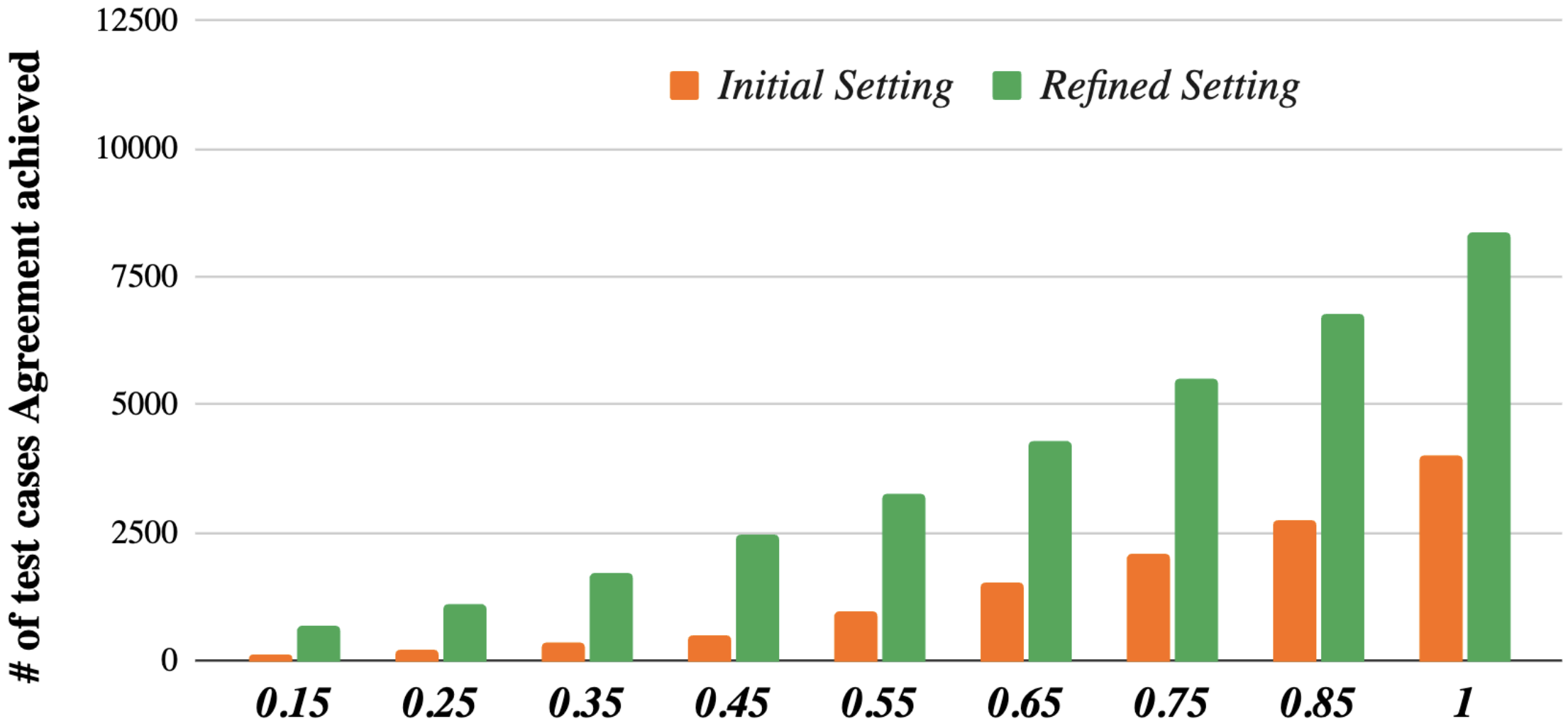}
\caption{\textbf{Agreement counts} (out of 12,700 test cases) across thresholds
$\alpha \in [0.15, 1.0]$ for initial and refined EZR setting.}
\vspace{-2ex}
\label{fig:sensitivity}
\end{figure}

\textbf{Performance stability} asks if repeated runs make consistent predictions, whatever their structure. We pass 100 randomly selected hold-out rows (fixed across treatments) through all 20 models. For each row, let $\sigma_{\text{models}}$ be the standard deviation of the 20 predictions and $\sigma_{\text{data}}$ the standard deviation of true $d2h$ across the dataset. The models \textit{agree} on that row when
\begin{equation}
    \sigma_{\text{models}} < \alpha \times \sigma_{\text{data}}
\label{eq:threshold}
\end{equation}
i.e., when disagreement is small against the background variation. Any such threshold constant is debatable, so we guard against it two ways. First, we report results primarily at $\alpha=0.35$. Second, RQ0 sweeps the full range $0.15 \le \alpha \le 1$, from a strict test ($\alpha=0.15$) to the most permissive one ($\alpha=1$, where models ``agree'' whenever their spread is merely smaller than the standard deviation of the entire dataset). As shown later (Figure~\ref{fig:sensitivity}), our conclusions hold across the sweep and are not artifacts of $\alpha$ choice.

Agreement aggregates two ways: the \textbf{agreement rate} (per dataset, the percent of the 100 rows where all 20 models agree) and the \textbf{total agreement count} (summed over all $127 \times 100 = 12{,}700$ test cases). Applying \S\ref{sec:statistics} to per-dataset agreement rates yields each treatment's \textit{stability win count}. Note that Figure~\ref{fig:std} reports stability in terms of standard deviation of optimization error (lower = more stable), while Figure~\ref{fig:sensitivity} reports agreement rates across test cases. Both measure performance instability but from complementary angles.

\begin{figure*}[t!]
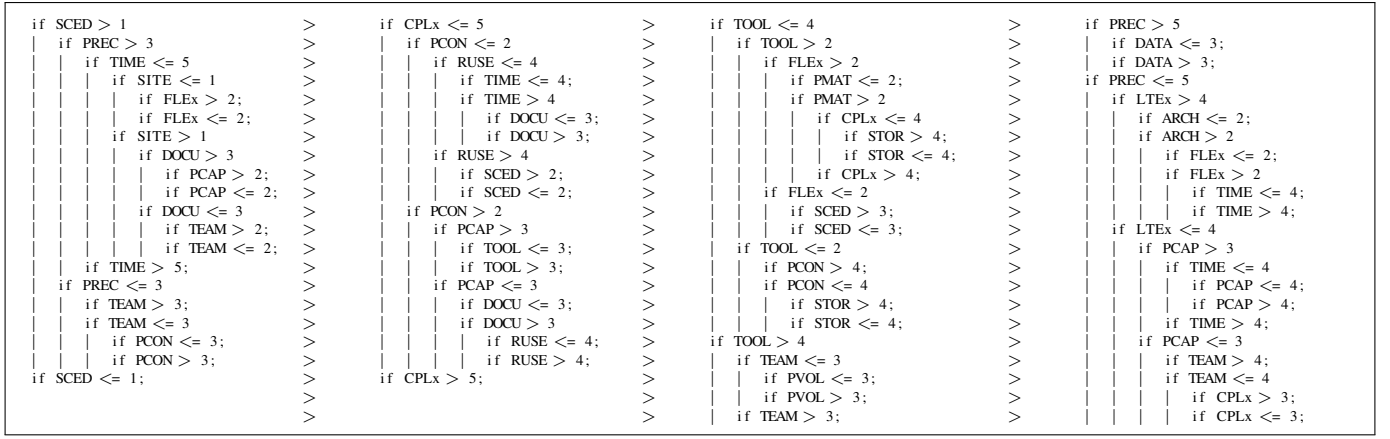

\centering
{\fontsize{5.6}{7}\selectfont
\begin{framed}
\vspace{-10pt}
\begin{lstlisting}
if SCED > 1                    >       if CPLx <= 5                  >      if TOOL <= 4                      >       if PREC > 5              
|  if PREC > 3                 >       |  if PCON <= 2               >      |  if TOOL > 2                    >       |  if DATA <= 3;         
|  |  if TIME <= 5             >       |  |  if RUSE <= 4            >      |  |  if FLEx > 2                 >       |  if DATA > 3;          
|  |  |  if SITE <= 1          >       |  |  |  if TIME <= 4;        >      |  |  |  if PMAT <= 2;            >       if PREC <= 5             
|  |  |  |  if FLEx > 2;       >       |  |  |  if TIME > 4          >      |  |  |  if PMAT > 2              >       |  if LTEx > 4           
|  |  |  |  if FLEx <= 2;      >       |  |  |  |  if DOCU <= 3;     >      |  |  |  |  if CPLx <= 4          >       |  |  if ARCH <= 2;      
|  |  |  if SITE > 1           >       |  |  |  |  if DOCU > 3;      >      |  |  |  |  |  if STOR > 4;       >       |  |  if ARCH > 2        
|  |  |  |  if DOCU > 3        >       |  |  if RUSE > 4             >      |  |  |  |  |  if STOR <= 4;      >       |  |  |  if FLEx <= 2;   
|  |  |  |  |  if PCAP > 2;    >       |  |  |  if SCED > 2;         >      |  |  |  |  if CPLx > 4;          >       |  |  |  if FLEx > 2     
|  |  |  |  |  if PCAP <= 2;   >       |  |  |  if SCED <= 2;        >      |  |  if FLEx <= 2                >       |  |  |  |  if TIME <= 4;
|  |  |  |  if DOCU <= 3       >       |  if PCON > 2                >      |  |  |  if SCED > 3;             >       |  |  |  |  if TIME > 4; 
|  |  |  |  |  if TEAM > 2;    >       |  |  if PCAP > 3             >      |  |  |  if SCED <= 3;            >       |  if LTEx <= 4          
|  |  |  |  |  if TEAM <= 2;   >       |  |  |  if TOOL <= 3;        >      |  if TOOL <= 2                   >       |  |  if PCAP > 3         
|  |  if TIME > 5;             >       |  |  |  if TOOL > 3;         >      |  |  if PCON > 4;                >       |  |  |  if TIME <= 4     
|  if PREC <= 3                >       |  |  if PCAP <= 3            >      |  |  if PCON <= 4                >       |  |  |  |  if PCAP <= 4; 
|  |  if TEAM > 3;             >       |  |  |  if DOCU <= 3;        >      |  |  |  if STOR > 4;             >       |  |  |  |  if PCAP > 4;  
|  |  if TEAM <= 3             >       |  |  |  if DOCU > 3          >      |  |  |  if STOR <= 4;            >       |  |  |  if TIME > 4;     
|  |  |  if PCON <= 3;         >       |  |  |  |  if RUSE <= 4;     >      if TOOL > 4                       >       |  |  if PCAP <= 3        
|  |  |  if PCON > 3;          >       |  |  |  |  if RUSE > 4;      >      |  if TEAM <= 3                   >       |  |  |  if TEAM > 4;     
if SCED <= 1;                  >       if CPLx > 5;                  >      |  |  if PVOL <= 3;               >       |  |  |  if TEAM <= 4     
                               >                                     >      |  |  if PVOL > 3;                >       |  |  |  |  if CPLx > 3;  
                               >                                     >      |  if TEAM > 3;                   >       |  |  |  |  if CPLx <= 3; 
\end{lstlisting}
\vspace{-12pt}
\end{framed}}
\vspace{-8pt}
\caption{Four trees trained on coc1000.csv (same data as Figure~\ref{fig:instability_example}) with RQ2-refined settings and different random seeds produce markedly different structures yet more consistent optimization recommendations compared to Figure~\ref{fig:instability_example}, illustrating that structural and performance instability are decoupled phenomena.}
\label{fig:instability_example_2}
\vspace{-10pt}
\end{figure*}

\subsection{Statistics}
\label{sec:statistics}
To compare treatments, we require statistical significance \emph{and} practical effect size, avoiding p-value-only pitfalls~\cite{kampenes2007systematic}. A K-S test~\cite{lilliefors1967kolmogorov} at $\alpha=0.05$ plus Cliff's delta $\delta \le 0.195$ (small-to-medium)~\cite{rosenthal1994parametric, sawilowsky2009new, macbeth2011cliff}. To rank many treatments we find the \textit{top set}: a variance-maximizing iterative partition~\cite{semenova2023path} peels away distinguishably worse treatments, retaining those statistically tied with the best.

\section{Results}
\label{sec:results}

\subsection{\textbf{RQ0 - The Problem:} How prevalent is performance instability in SE optimization, and what are its consequences for practitioners?}
\label{sec:RQ0_results}

Using \S\ref{sec:stab_measure}, we train 20 trees per dataset under two settings (initial EZR defaults, the refined settings later recommended by RQ2) and count test-case agreements. At $\alpha=0.35$, the 20 trees agree on just 364 (Initial setting) and 1,740 (Refined setting) of 12,700 test cases: consistency rates of \textbf{2.9\%} and \textbf{13.7\%}. Figure~\ref{fig:sensitivity} shows results sweeping the $\alpha$ threshold for both our baseline system (in red) and the refined system (in green) recommended later in this paper. This figure confirms that even at $\alpha=1$, agreement reaches only 4,020 and 8,360 cases (31.7\% and 65.8\%), and the refined configuration remains better.

The practical consequence is that even under the most permissive threshold, and even after refinement, nearly half of all test cases fail to agree. Two runs of the same tool on the same data will usually return different predictions, undermining a team's ability to reliably allocate testing budget, prioritize high-risk modules, or select near-optimal configurations.
\begin{keybox}
\textbf{RQ0 answer.} Performance instability is the norm, not the exception. Under default settings, repeated runs agree on 2.9\% of test cases, and no choice of threshold rescues consistency. Practitioners may not be able to trust the recommendation of any single run.
\end{keybox}

\begin{figure*}[!t]
\centering
\includegraphics[width=\textwidth]{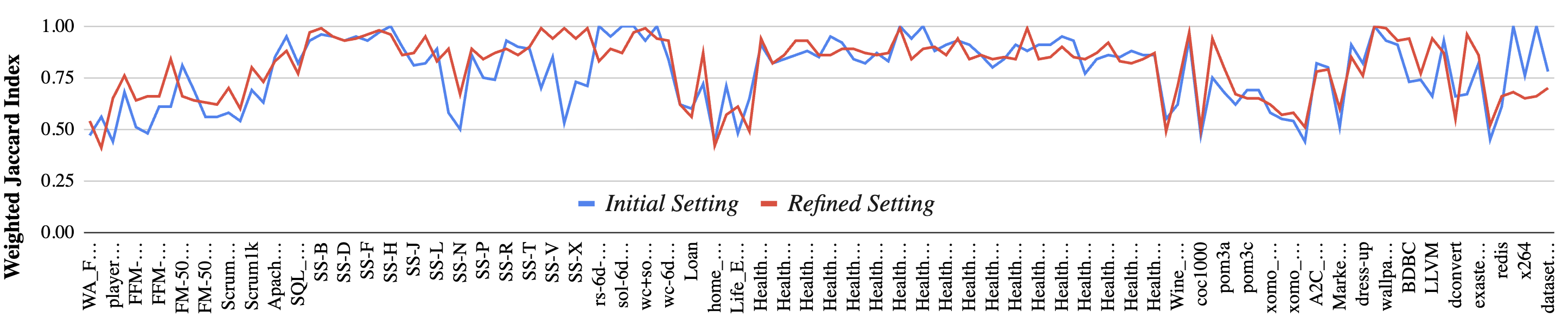}
\caption{\textbf{Structural similarity} across 20 trees per dataset under initial and refined EZR settings. Refined settings greatly improve performance stability (Fig.~\ref{fig:rq2_final_comparison}) while producing little change in structural similarity, suggesting the two are decoupled.}
\vspace{-2ex}
\label{fig:jaccard}
\end{figure*}

\subsection{\textbf{RQ1 - The Nature of the Problem:} How prevalent is structural instability, and is it the same phenomenon as performance instability?}
\label{sec:RQ1_results}

Using the weighted Jaccard measure of \S\ref{sec:stab_measure}, Figure~\ref{fig:jaccard} compares the feature sets of the 20 trees per dataset under the {\em initial} and {\em refined} configurations (recall from \S\ref{sec:ezr} that the refined settings, derived in RQ2, are those that most improve performance stability).

\begin{figure*}[!b]
\vspace{-10pt}
  \centering
  \includegraphics[width=\linewidth]{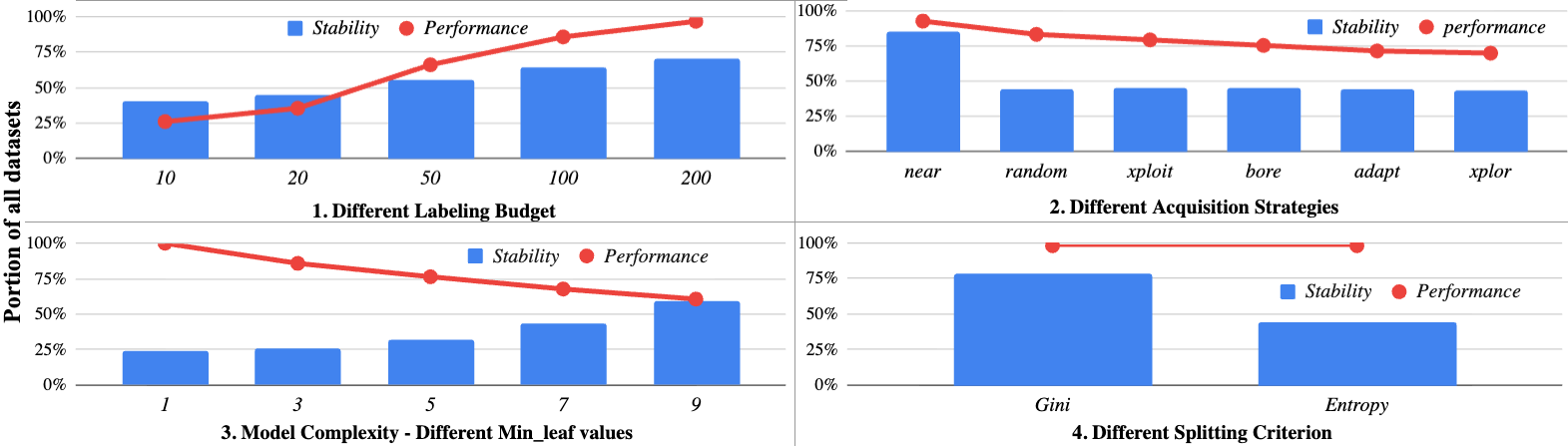}
  \caption{Summary of four instability-cause experiments. Values indicate the percentage of datasets where each treatment achieved the highest statistical rank. Each panel reports stability (blue bars) and performance (red line or bars) across treatments.}
  \vspace{-20pt}
  \label{fig:rq2_combined}
\end{figure*}

Two features of Figure~\ref{fig:jaccard} are noteworthy. {\bf Firstly}, structural and performance stability are decoupled problems. The two curves, remaining closely aligned, show that the refined settings that improve prediction agreement more than sixfold (RQ0) produce trees that are structurally as different as the defaults. Figure~\ref{fig:instability_example_2} showed the effect on one dataset. Four refined-setting trees still differ markedly in features, thresholds, and topology. One practical corollary: the features of any single tree should not be over-interpreted as {\em the} explanation of a recommendation, since other runs select other features while recommending the same things. {\bf Secondly}, structural instability is not the instability that should be repaired, for two reasons. First, it is largely irreducible: the Rashomon effect (\S\ref{sec:rashomon}) guarantees many feature sets fit equally well, which is why even our most stabilizing settings barely move structural similarity. Second, it may not be what hurts practitioners. The decoupling shown above means trees can differ in structure yet agree on their recommendations, so it is performance agreement, not feature overlap, that governs whether a single run can be trusted.
\begin{keybox}
\textbf{RQ1 answer.} Structural and performance instability are distinct. Settings that sharply improve prediction agreement leave tree structure about as varied as before. We target performance instability because it is the fixable one that governs trust. Structural instability is largely irreducible (Rashomon) and, being decoupled from recommendations, need not be fixed.
\end{keybox}

\subsection{\textbf{RQ2 - The Fixes:} Which factors drive performance instability, and what configuration changes reduce it without sacrificing optimization quality?}
\label{sec:RQ2_results}

Figure~\ref{fig:rq2_combined} explores parts of Table~\ref{tab:instability_causes} one at a time. The main finding here is that stability and performance respond to different levers, sometimes independently, sometimes in opposition, and no single cause dominates.

\textbf{Labeling budget} can improve both performance and structural metrics. As seen at the top-left of Figure~\ref{fig:rq2_combined}, increasing the label budget raises the fraction of datasets where the treatment is statistically top-ranked for stability from under 40\% to around 70\%. However, for two reasons, chasing more labels is not our preferred fix for instability since:
\begin{itemize}
\item Performance gains saturate. As seen top-left of Figure~\ref{fig:rq2_combined}, improvements beyond 50 labels are relatively modest, and a performance ceiling of 100\% is reached at 200 labels.
\item Other factors do as much for free. As shown below, the acquisition strategy buys comparable stability gains with no extra labels.
\end{itemize}

The \textbf{acquisition strategy} shows a much stronger effect.  Of the six strategies. \emph{Near} (label the candidate closest to the current best centroid) lifts stability top-rant rate from under 50\% to over 80\%. Note that adopting the \emph{near} comes at almost no cost (no extra labels, no change of objective, no loss of recommendation quality).

\begin{figure*}[t]
  \centering
  \includegraphics[width=\textwidth]{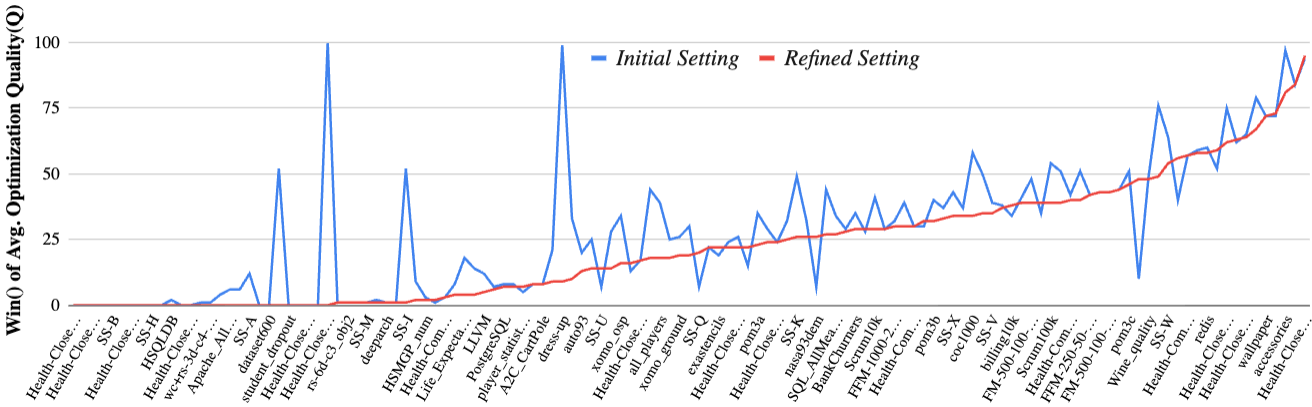}
  \caption{\textbf{Performance results} (\textit{win} of optimization quality $Q$, defined in Equation~\ref{eq:q})   across 127 datasets, measured as average deviation from the referenced optimum (lower is better). Datasets sorted by initial performance. Refined settings are statistically top-ranked on 119 datasets versus 74 for the initial configuration.  Note that, with our methods (the red line), optimization quality is nearly always improved, sometimes by very large amounts (e.g., see left-hand-side).}
  \label{fig:rq2_final_comparison}
  \vspace{-10pt}
\end{figure*}

\textbf{Model complexity} (minimum leaf size) trades three ways. Deeper trees find better configurations but are less stable (finer partitions are more sensitive to sampling noise) and harder for practitioners to read, undermining the very point of human-readable rules. \textit{min\_leaf} = 3 best balances all three.

\textbf{Splitting criterion} is a smaller but free win. Entropy's $\log(p)$ amplifies tiny perturbations in rare-event probabilities into large swings in split selection, precisely under the tight budgets where SE optimizers run (\S\ref{sec:causes}). Gini matches entropy's performance with clearly better stability. A zero-cost change, applicable immediately.

\begin{table}[!t]
\centering
\small
\begin{tabular}{lll}
Factor & Default & Recommended \\
\hline
Labeling budget & 20 & 50 \\
Acquisition strategy & Xploit & near \\
Min leaf size & 2 & 3 \\
Splitting criterion & Entropy & Gini \\

\end{tabular}
\setlength{\abovecaptionskip}{4pt}
\setlength{\belowcaptionskip}{-15pt}
\caption{Suggested EZR settings from RQ2 experiments.}
\label{tab:config_recommendation}
\end{table}

Varying one factor at a time cannot reveal interactions, so the four recommended settings (Table~\ref{tab:config_recommendation}) must also be evaluated as a single \emph{refined} configuration. Figures~\ref{fig:rq2_final_comparison} and~\ref{fig:std} provide that validation. Across all 127 datasets, the refined configuration reduces optimization-quality standard deviation by 22\% on average (Figure~\ref{fig:std}) while achieving statistically top-ranked optimization quality (\S\ref{sec:statistics}) on 119 datasets, compared to 74 for the default configuration (Figure~\ref{fig:rq2_final_comparison}). Thus, the improvements observed for individual factors compound. Refined setting improves both stability and optimization quality.

\begin{keybox}
\textbf{RQ2 answer.} Labeling budget, acquisition strategy, model complexity, and splitting criterion each drive performance instability, via different trade-offs. The refined configuration  Table~\ref{tab:config_recommendation}     (50 labels, \emph{near}, min\_leaf=3, Gini) improves stability (22\% reduction in prediction-error std) and quality simultaneously, achieving top-ranked optimization on 119 datasets versus 74 for initial settings.
\end{keybox}

\subsection{\textbf{RQ3 - The Limits:} Is the instability remaining after RQ2 the fault of the learner, or of
the data?}
\label{sec:Causal-Experiments} 
The previous section focused on the learner, tuning internal choices such as budget, acquisition strategy, leaf size, and splitting criterion. Here we shift focus to the data. The question is whether instability is driven primarily by spurious correlations among features or by heterogeneous subpopulations hidden within the rows. We test this hypothesis in two ways: by augmenting EZR with causal knowledge (confounder removal), and by replacing trees with stable-by-design clusterers~\cite{breiman1996heuristics} that model local regions rather than a single global structure. If data-side effects dominate, these interventions should deliver gains far larger than those seen in RQ2. As shown below, they do not.

\begin{figure}[!t]
\centering
\includegraphics[width=\columnwidth]{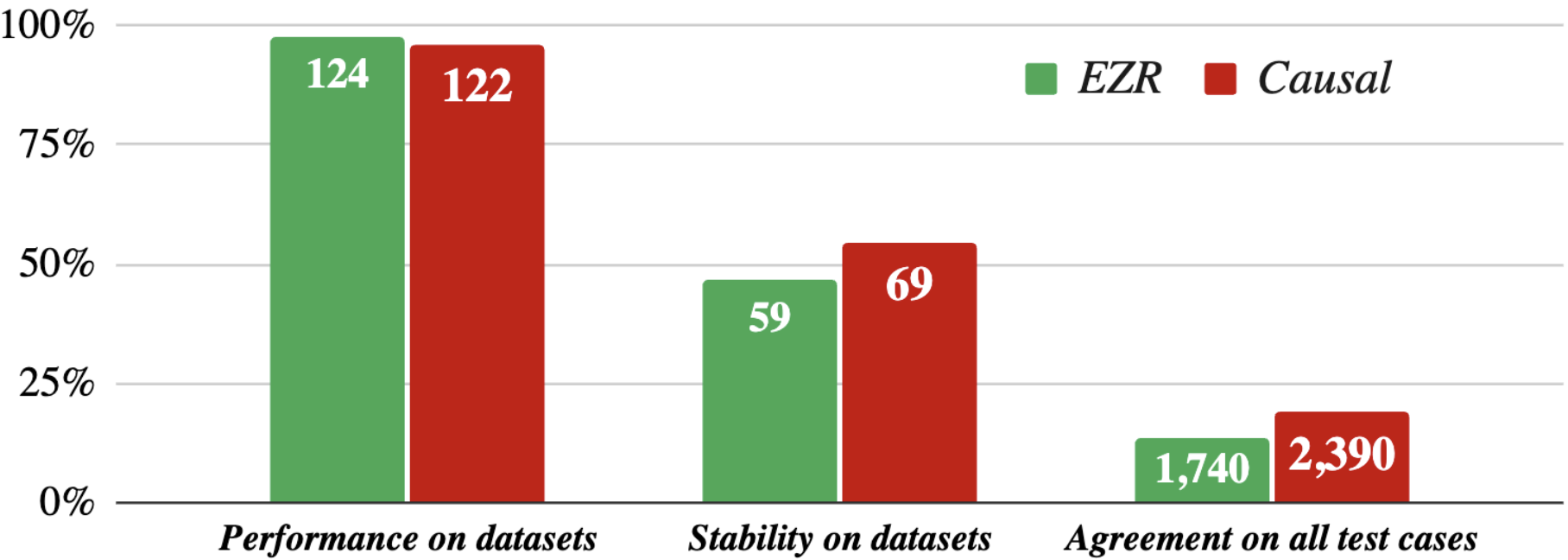}
\caption{\textbf{Performance} (dataset statistical wins), \textbf{stability} (dataset statistical wins), and \textbf{agreement} (test-case count out of 12,700) for \textbf{EZR and causal pipeline} across all datasets.}
\label{fig:RQ3-results}
\vspace{-15pt}
\end{figure}

\textbf{Causal-inspired reasoning:} comparing the causal pipeline (confounder filter + gain-ratio splits, \S\ref{sec:causes}) against refined EZR, performance is statistically indistinguishable (Figure~\ref{fig:RQ3-results}: 124 vs.\ 122 dataset wins of 127), so filtering removes no useful signal. But its stability gains are confined to some datasets. Causal pipeline wins on 10 more datasets (59 vs.\ 69) and, also collects more total agreements (1,740 vs. 2,390), although this difference is not significant. In short, causal pipeline helps only where confounded features drive instability, and elsewhere noise, small samples, or large Rashomon sets (\S II.B) dominate (most datasets).

\begin{figure}[!t]
\centering
\includegraphics[width=\columnwidth]{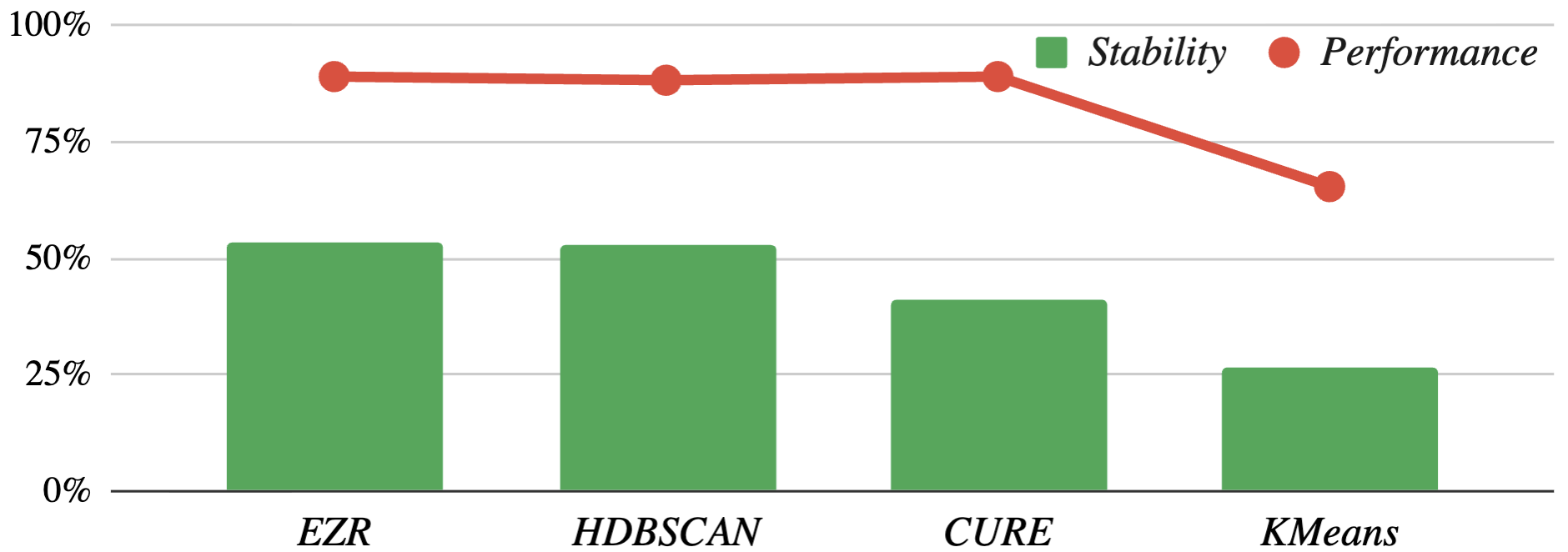}
\caption{Comparing Performance and stability of EZR with \textbf{clustering methods}. Green bars and red lines show the stability and performance scores across 127 datasets.}
\label{fig:RQ3-2-results}
\vspace{-10pt}
\end{figure}

\textbf{Data-inherent limits:} if instability were something a better learner could remove, the stablest learners we know should remove it. Clustering methods do lead clearly (Figure~\ref{fig:RQ3-2-results}). But even the best of them agree on less than 51\% of cases. Switching to the stablest model family buys at most half the benchmark, evidence (not proof) that the floor is set by the data itself: noise, tight labeling budgets, proxy objectives, and large Rashomon sets.
 
\begin{keybox}
\textbf{RQ3 answer.} Instability is not fully curable by changing the learner. External interventions add little. Causal pipeline helps only on some datasets, and even the stablest clusterers agree on under 51\% of cases. The residue appears data-inherent, a limit shared by all methods, to be measured and reported, not blamed on the tool.
\end{keybox}

\section{Threats to Validity}
\label{valid}
\textbf{Sampling bias.} No single paper can investigate every problem in
software analytics. Hence, when we say ``our methods tame instability,'' we add the caveat: we have shown this only for the examples studied here. That said, our examples were many and varied: 127 optimization tasks, written by many authors, from many leading venues (\S\ref{sec:exp-datasets}), all processed by a learner with recent state-of-the-art credentials
(\S\ref{sec:ezr}).

\textbf{External validity.} We studied one family of learners (tree-based active learners) plus three clusterers. Our specific refined settings (Table~\ref{tab:config_recommendation}) may not transfer to, say, neural optimizers. That said, RQ3 suggests the data-inherent stability floor is shared across very different model families, though that is evidence, not proof, and checking other learners is future work.

\textbf{Construct validity.} Instability is not a quantity with one agreed operationalization, and this paper measured it just two ways: structural instability via a weighted Jaccard over tree feature sets, and performance instability via cross-run agreement on predictions (\S\ref{sec:stab_measure}). Other instruments (e.g. tree edit distance, rank correlation of recommendations, prediction-interval overlap) might score the same runs differently, and we cannot rule out that our conclusions are tied to our measures. Three things mitigate this concern. First, both measures have precedent: weighted Jaccard follows the feature-stability literature~\cite{10.1007/s10115-006-0040-8}, and our agreement threshold is grounded in standard effect-size practice~\cite{rosenthal1994parametric, sawilowsky2009new, macbeth2011cliff}. Second, the threshold constant in our agreement test was swept across its full range (Figure~\ref{fig:sensitivity}) and our conclusions held at every setting. Third, our labeling budgets were capped at an empirically observed performance ceiling (200 labels \S\ref{sec:exp-datasets}), not an arbitrary choice. Still, replication with other instability measures would be a valuable check on this work.

\section{Conclusion}
\label{sec:conclusion}

Performance instability is not a minor reproducibility nuisance, and it is a threat to trust in SE optimization. Under default settings, repeated runs agreed on only 2.9\% of 12{,}700 test cases(RQ0). Refined settings (Table~\ref{tab:config_recommendation}) increased agreement by 4.8$\times$, reduced the standard deviation of optimization error by 22\%, and improved optimization quality, achieving the statistical top rank on 119 of 127 datasets, up from 74 for the initial configuration(RQ2). These gains came from simple learner-side choices: using \emph{near} instead of greedy-exploit acquisition, Gini instead of entropy, \emph{min\_leaf}=3, and a modest label budget of 50. 

But the larger lesson is not that instability disappears. It does not. Even after refinement, agreement reaches only 13.7\% at our main threshold, and even the stablest learners agree on fewer than 51\% of test cases(RQ3). Structural instability also remains widespread, confirming that different model structures can still support similar recommendations(RQ1). Thus, the goal should not be to force repeated runs into the same tree. The goal should be to know when to trust their recommendations.

These findings imply three recommendations. First, SE optimization tools should reconsider unstable defaults, more specifically, entropy splitting and greedy-exploit acquisition. Second, empirical SE papers should report agreement rates alongside optimization results, since mean performance alone hides cross-run disagreement. Third, researchers should avoid two shortcuts. First, chasing structural stability for its own sake, since Rashomon makes it unlikely, and RQ1 shows it is unnecessary; And second, do not expect a silver bullet. Instability has multiple causes, and effective fixes compound rather than replace one another. Stability should be a standard evaluation axis. Without it, optimization result is incomplete.

As to future work, we recommend (a) treating stability, performance, and explainability as three objectives on a shared Pareto front; (b) attacking the data-inherent ceiling upstream, via better measurement and labeling practices; (c) repeating this analysis on other interpretable learners (rule lists, Bayesian models, association rule miners); and (d) testing causal augmentation in domains rich in organizational confounders, where RQ3 suggests it should shine.

\balance
\bibliographystyle{IEEEtran}
\bibliography{references}

\end{document}